\documentclass[11pt]{article}
\usepackage{epsfig}
\usepackage{graphicx}
\usepackage{epsfig}

\begin{document}

\title{\bf Finding the reconstructions of semiconductor surfaces via a genetic algorithm}

\author{F.~C.~Chuang$^{1}$, C.~V.~Ciobanu$^{2,3}$\footnote{Corresponding author}, V.~B.~Shenoy$^2$, C.~Z.~Wang$^1$, and K.~M.~Ho$^1$ \\
$^1$Ames Laboratory -- U.S. Department of Energy and \\Department
of Physics, Iowa State University, Ames, Iowa 50011, USA  \\
$^2$Division of Engineering, Brown University, Providence, Rhode
Island 02912, USA \\
$^3$Division of Engineering, Colorado School of Mines, Golden,
Colorado 80401, USA}

\maketitle

\bigskip
\bigskip
\bigskip
\bigskip

\begin{abstract}
In this article we show that the reconstructions of semiconductor
surfaces can be determined using a genetic procedure. Coupled with
highly optimized interatomic potentials, the present approach
represents an efficient tool for finding and sorting good
structural candidates for further electronic structure
calculations and comparison with scanning tunnelling microscope
(STM) images. We illustrate the method for the case of Si(105),
and build a database of structures that includes the previously
found low-energy models, as well as a number of novel
configurations.
\end{abstract}

\maketitle \newpage

The determination of atomic structure of crystalline surfaces is a
long-standing problem in surface science. Despite major progress
brought by experimental techniques such as scanning tunnelling
microscopy (STM) and advanced theoretical methods for treating the
electronic and ionic motion, the commonly used procedures for
finding the atomic structure of surfaces still rely to a large
extent on one's intuition in interpreting STM images. While these
procedures have proven successful for determining the atomic
configuration of many low-index surfaces [e.g., Si(001) and
Si(111)], in the case of high-index surfaces their usefulness is
limited because the number of good structural models for
high-index surfaces is rather large, and may not be exhausted
heuristically. An illustrative example is Si(5 5 12), whose
structure has been the subject of intense dispute
\cite{5512-ranke, 5512suzuki-both, 5512takeguchi-both, 5512new}
since the publication of the first atomic model proposed for this
surface \cite{5512old}. While the structure of Si(5 5 12) may
still be an open problem, there are other stable surfaces of
silicon such as (113) \cite{113briefreview} and (105) that
required a long time for their structures to be revealed; in the
present work we focus on Si(105), and show that there is a large
number of low-energy models for this surface.

The high-index surfaces attract a great deal of scientific and
technological interest since they can serve as natural and
inexpensive templates for the fabrication of low-dimensional
nanoscale structures. Knowledge about the template surface can
lead to new ways of engineering the morphological and physical
properties of these nanostructures. The main technique for
investigating atomic-scale features of surfaces is STM, although,
as pointed out in a recent review, STM alone is only able to
provide "a range of speculative structural models which are
increasingly regarded as solved surface structures"
\cite{woodruff}. With few exceptions that concern low-index
metallic surfaces \cite{kress, carter}, the role of theoretical
methods for structural optimization of surfaces has been largely
reduced to the relaxation of these speculative models. However,
the publication of numerous studies that report different
structures for a given high-index silicon surface (see, e.g.,
\cite{5512-ranke, 5512suzuki-both, 5512takeguchi-both, 5512new,
5512old}) indicates a need to develop methodologies capable of
actually searching for the atomic structure in a way that does not
predominantly rely on the heuristic reasoning associated with
interpreting STM data. Recently, it was shown that
parallel-tempering Monte Carlo (PTMC) simulations combined with an
exponential cooling schedule can successfully address the problem
of finding the reconstructions of high-index silicon surfaces
\cite{ptmc}. The PTMC simulations, however, have a broader scope,
as they are used to perform a thorough thermodynamic sampling of
the surface systems under study. Given their scope, such
calculations \cite{ptmc} are very demanding, usually requiring
several tens of processors that run canonical simulations at
different temperatures and exchange configurations in order to
drive the low-temperature replicas into the ground state. If we
focus only on finding the reconstructions at zero Kelvin (which
can be representative for crystal surfaces in the low-temperature
regimes achieved in laboratory conditions), it is then justified
to explore alternative methods for finding the structure of
high-index surfaces.

In this letter, we address the problem of surface structure
determination at zero Kelvin, and report a genetically-based
strategy for finding the reconstructions of elemental
semiconductor surfaces. Our choice for developing this genetic
algorithm (GA) was motivated by its successful application for the
structural optimization of atomic clusters \cite{ga-prl,
ga-nature}. Like the previous study \cite{ptmc}, the present
algorithm also circumvents the intuitive process when proposing
candidate models for a given high-index surface. Except for the
periodic vectors of the surface unit cell [which can be determined
from scanning tunnelling microscopy (STM) or from low-energy
electron diffraction (LEED) measurements], no other experimental
input is necessary. An advantage of the present approach over most
of the previous methodologies used for structural optimization is
that the number of atoms involved in the reconstruction, as well
as their most favorable bonding topology, can be found within the
same genetic search. Since the interactions are modelled by
empirical potentials, it is generally desirable to check the
relative stability of different model structures using
higher-level calculations based on density functional theory. Here
we test the genetic procedure on the (105) surface, which, at
least in conditions of compressive strain, is known to have a
single-height rebonded step structure SR \cite{jap-prl, italy-prl,
apl, susc105, montalenti}. The PTMC study \cite{ptmc} indicates
that the SR structure is the lowest surface energy even in the
absence of strain, although there are several other
reconstructions with very similar surface energies. It is
interesting to note that the number of reported reconstructions
for the (105) orientation has increased very rapidly from two
(models SU and SR \cite{jap-prl, italy-prl, apl}), to a total of
fourteen reported in Refs. \cite{ptmc, susc105}. While the set of
known reconstructions has expanded, the most favorable structure
has remained the SR model --in contrast with the first reported
model \cite{mo}, but in agreement with more recent studies
\cite{jap-prl, italy-prl, apl}.

Before describing the algorithm, we pause to briefly discuss the
computational details. The simulation cell has a single-face slab
geometry with periodic boundary conditions applied in the plane of
the surface, and no periodicity in the direction normal to it. The
top atoms corresponding to a depth $d=5$\AA \ (measured from the
position of the highest atom) are shuffled via a set of genetic
operations described below. In order to properly account for the
surface stress, the atoms in a thicker zone of 15--20\AA \ are
allowed to relax to a local minimum of the potential energy after
each genetic operation. In the present work, we test the algorithm
for the case of Si(105); the surface slab is made of four bulk
unit cells of dimensions\footnote{The lengths of the periodic
vectors that correspond to the bulk unit cell are determined from
analytic geometry calculations and knowledge of the crystal
structure.} $a\sqrt{6.5}\times a \times a\sqrt{6.5}$ ($a=5.431$\AA
\ is the bulk lattice constant of Si), stacked two by two along
the $[010]$ and [105] directions. In terms of atomic interactions,
we have used the highly-optimized empirical model developed by
Lenosky {\em et al.} \cite{hoep}, which was found to have superior
transferability to the diverse bonding environments present on
high-index silicon surfaces \cite{ptmc}.

The optimization procedure developed here is based on the idea of
evolutionary approach in which the members of a generation (pool
of models for the surface) mate with the goal of producing the
best specimens, i.e. lowest energy reconstructions. "Generation
zero" is a pool of $p$ different structures obtained by
randomizing the positions of the topmost atoms (thickness $d$),
and by subsequently relaxing the simulation slabs through a
conjugate-gradient procedure. The evolution from a generation to
the next one takes place by mating, which is achieved by
subjecting two randomly picked structures from the pool to a
certain operation ${\cal O}$:(A,B)$\longrightarrow$C. Before
defining this operation, we describe how the survival of the
fittest is implemented. In each generation, a number of $m$ mating
operations are performed. The resulting $m$ children are relaxed
and considered for the possible inclusion in the pool based on
their surface energy. If there exists at least one candidate in
the pool that has a higher surface energy than that of the child
considered, then the child structure is included in the pool. Upon
inclusion of the child, the structure with the highest surface
energy is discarded in order to preserve the total population $p$.
As described, the algorithm favors the crowding of the ecology
with identical metastable configurations, which slows down the
evolution towards the global minimum. To avoid the duplication of
members, we retain a new structure only if its surface energy
differs by more than $\delta$ when compared to the surface energy
of any of the current members $p$ of the pool. We also consider a
criterion based on atomic displacements to account for the
(theoretically possible) situation in which two structures have
equal energy but different topologies: two models are considered
structurally different if the relative displacement of at least
one pair of corresponding atoms is greater than $\epsilon$.
Relevant values for the parameters of the algorithm are $10\leq p
\leq 40$, $m=10$, $d=5$\AA \ , $\delta=10^{-5}$meV/\AA$^2$, and
$\epsilon =0.2$\AA .

\begin{figure}
\begin{center}
\epsfig{file=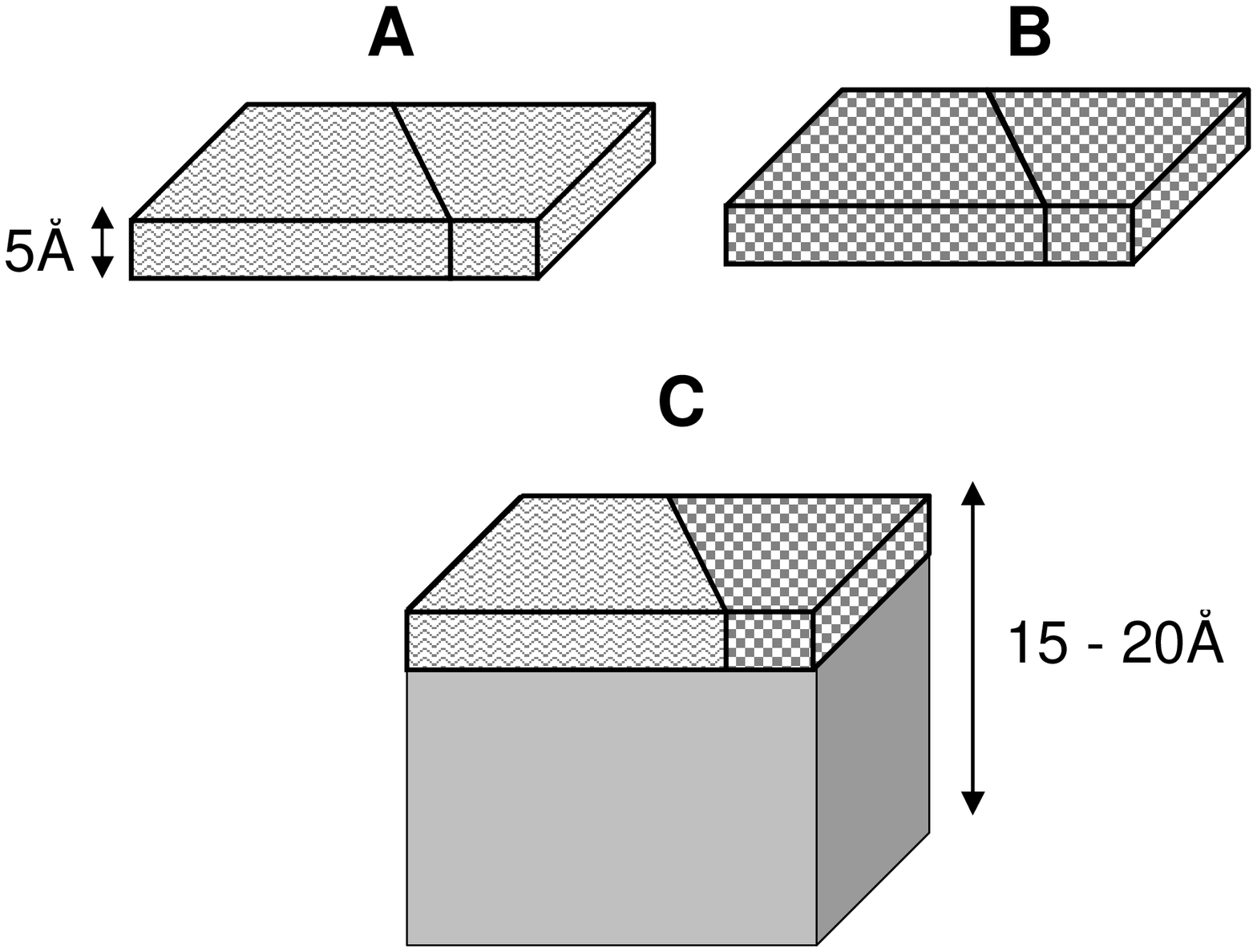,width=10.0cm} %
\caption{The mating operation ${\cal O}:$ (A,B)~$\rightarrow$~C.
From two candidate surface structures A and B, the upper portions
(5\AA \ -thick) are separated and sectioned by the same arbitrary
plane oriented perpendicular to the surface. A new slab C is
created by combining the part of A that lies to the left of the
cutting plane and the part of slab B lying to the right of that
plane. C is placed on a thicker slab, and the resulting structure
is relaxed before considering its inclusion in the pool of
candidate reconstructions. }\label{mating}
\end{center}
\end{figure}

We now describe the mating operation, which produces a child
structure from two parent configurations as follows (refer to
Fig.~\ref{mating}). The topmost parts of the parent models A and B
(thickness $d$) are separated from the underlying bulk and
sectioned by an arbitrary plane perpendicular to the surface. The
(upper part of the) child structure C is created by combining the
part of A that lies to the left of the cutting plane and the part
of slab B lying to the right of that plane: the assembly is placed
on a thicker slab, and the resulting structure C is relaxed. We
have found that the algorithm is more efficient when the cutting
plane is not constrained to pass through the center of the surface
unit cell, and also when that plane is not too close to the cell
boundaries. Therefore, we pick the cutting plane such that it
passes through a random point situated within a rectangle centered
inside the unit cell; numerical experimentation has shown that the
algorithm performs very well if the area of that rectangle is
about 80\% of the area of the surface cell. We have developed two
versions of the algorithm. In the first version, the number of
atoms $n$ is kept the same for every member of the pool by
automatically rejecting child structures that have different
numbers of atoms from their parents (mutants). In the second
version of the algorithm, this restriction is not enforced, i.e.
mutants are allowed to be part of the pool. As we shall see, the
procedure is able to select the correct number of atoms for the
ground state reconstruction without any increase over the
computational effort required for one single constant-$n$ run.

\begin{figure}
\begin{center}
 \includegraphics[width=7.50cm]{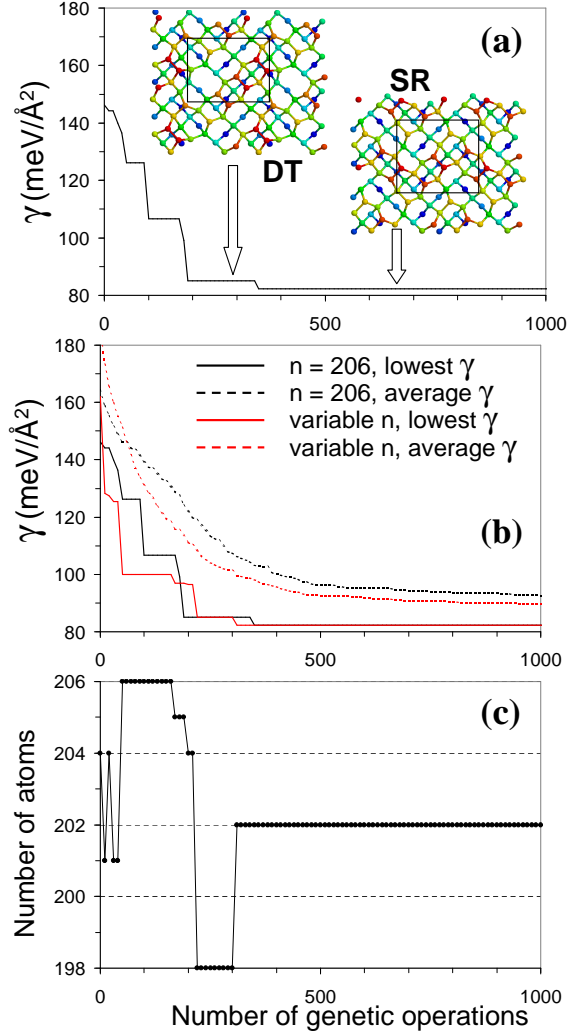}
\caption{(a) Surface energy $\gamma$ of the most stable Si(105)
candidate from a pool of $p=30$ structures (206-atom slabs with
dimensions $a\sqrt{6.5}\times 2a$), plotted as a function of the
number of mating operations. The genetic algorithm quickly
retrieves the DT structure \cite{ptmc} and the global minimum
structure, SR. The insets show top views (i.e. along the
$[\overline{1}0\overline{5}]$ direction) of the DT and SR models;
atoms are rainbow-colored according to their coordinate along the
[105] direction (red being the highest), and the rectangles show
the surface unit cells. (b) Comparison between runs with variable
number ($198\leq n \leq 210$) of atoms (red lines) and constant
$n,\ n=206$ (black lines). The lowest (average) surface energies
are shown as solid (dashed) lines. (c) The variation of the number
of atoms of the lowest energy configuration shows that the fittest
member of the pool eventually selects a value of $n$ that is
compatible with the global minimum structure SR.
}%
\label{DTSR-ga-abc}%
\end{center}
\end{figure}

The results for a Si(105) slab with 206 atoms (constant $n$) are
summarized in Fig.~\ref{DTSR-ga-abc}(a), which shows the surface
energy of the most stable member of a pool of $p=30$ candidates as
a function of the number of genetic operations. The lowest surface
energy starts at a very high value because the members of early
generations have random surface topologies. We find that in less
than 200 mating operations the most favorable structure in the
pool is already reconstructed, i.e. each atom at the surface has
at most one dangling (unsatisfied) bond. Furthermore, the density
of dangling bonds (db) per unit area is the smallest possible for
the Si(105) surface: the structure obtained is a double-step model
termed DT \cite{ptmc} that has 4 db/$a^2\sqrt{6.5}$. The
single-height rebonded structure SR \cite{jap-prl, italy-prl, apl}
is retrieved in less than 400 mating operations. The SR model is
in fact the global minimum for Si(105), as found recently in an
exhaustive PTMC search \cite{ptmc}. We have verified this finding
by performing constant-$n$ GA runs for a set of consecutive
numbers of atoms, $n=206,205,204,$ and 203.

However, we take a further step in that we seamlessly integrate
the search for the correct number of atoms within the search for
the lowest-energy reconstruction: we achieve this by allowing
energetically fit mutants to survive during the evolution, instead
of restricting the number of atoms to be constant across the pool.
The results from a GA run with variable $n$ are shown in
Fig.~\ref{DTSR-ga-abc}(b), in comparison with an $n=206$ run. We
notice that the variable-$n$ evolution displays a faster drop in
the lowest surface energy, as well as in the average energy across
the pool. For performance testing purposes, we started the
variable-$n$ run with all the candidates having an unfavorable
number of particles, $n=204$: nevertheless, the most stable member
in the pool predominantly selected a number of atoms that allows
for the SR topology, i.e. $n=$198, 202, 206 (refer to
Fig.~\ref{DTSR-ga-abc}(c)). While we find no significant
difference in the computational effort required by the two
different evolutions, the variable-$n$ and the constant-$n$
($n=206$), the former is to be preferred: since the surface energy
of a Si(105) slab is a periodic function of the number of atoms
\cite{ptmc} with a period of $\Delta n=4 $, the variable-$n$ run
is ultimately four times faster than the sequential constant-$n$
searches. The results from the sequential runs are summarized in
Table~\ref{tabelul}, which shows the surface energies of twenty
structures from runs with $n=$206, 205, 204, and 203.

\begin{figure}
\begin{center}
 \includegraphics[width=10.0cm]{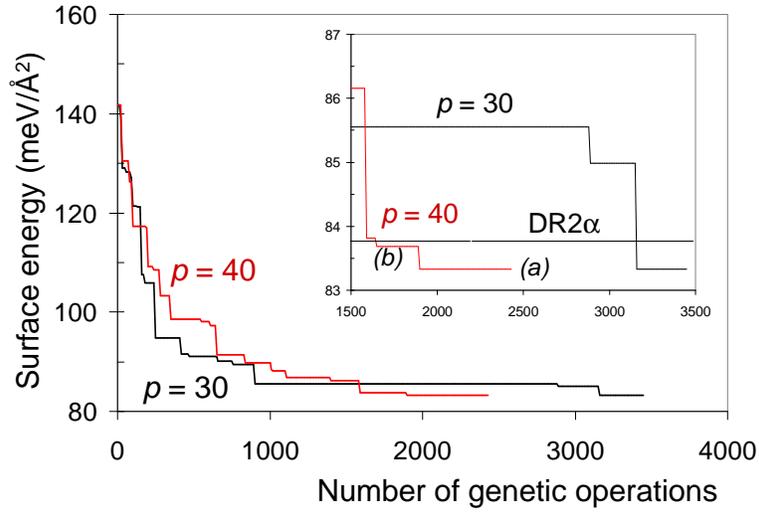}
\caption{Lowest surface energy $\gamma$ for pools of $p=30$ and
$p=40$ Si(105) candidate models. The simulation slab has a
periodic cell of $2a\sqrt{6.5}\times 2a$ and contains $n=406$
atoms, of which the highest-lying 70 atoms are subject to mating
operations. As seen in the inset, the procedure finds two
structures ($(a)$ and $(b)$), that are slightly more stable than
the DR2$\alpha$ model reported in Ref. \cite{ptmc}; these
configurations are shown in
Fig.~\ref{plateaus406}, along with DR2$\alpha$.}%
\label{fig:406-ga}
\end{center}
\end{figure}

\begin{figure}
\begin{center}
 \includegraphics[width=10.0cm]{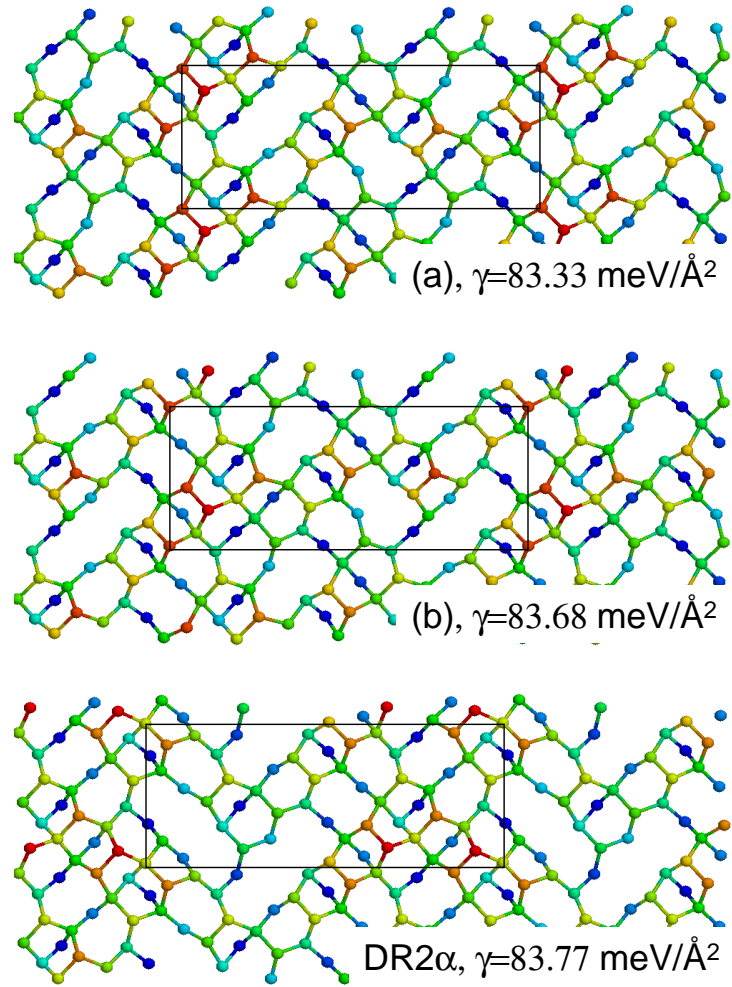}
\caption{Low-energy structures of the $2a\sqrt{6.5}\times 2a $,
406-atom surface unit cells (rectangles) for Si(105). The surface
energies $\gamma$ are indicated next to the corresponding model
labels. Atoms are rainbow-colored according to their height, as
described in Fig.~\ref{DTSR-ga-abc}.
}%
\label{plateaus406}
\end{center}
\end{figure}

Motivated by recent experimental work \cite{china-Si105} that
suggests the presence of a structure with large periodic length in
the [$50\overline{1}$] direction,  we have also performed a GA
search for configurations with larger surface unit cells
($2a\sqrt{6.5}\times 2a$), with $n=406$ atoms. A low-energy (105)
reconstruction of this size (termed DR2) was reported in our
earlier work \cite{susc105}, where we showed that step-edge
rebonding  lowers the surface energy below that of the DU1 model
proposed in Ref.\cite{china-Si105}. Upon annealing, the DR2 model
can evolve into three structures with lower surface energies,
DR2$\gamma$, DR2$\beta$ and DR2$\alpha$ \cite{ptmc}. Using the GA
technique described above, we find structures that are even more
favorable than DR2$\alpha$ (refer to Fig.~\ref{fig:406-ga} and
\ref{plateaus406}). The results displayed in Fig.~\ref{fig:406-ga}
indicate that the efficiency of the algorithm improves upon
increasing the number of candidates in the pool, from $p=30$ to
$p=40$. In both cases, the evolution retrieves in the same lowest
energy structure with $\gamma=83.33$ meV/\AA$^2$, which is nearly
degenerate with DR2$\alpha$, as the surface energy difference is
only $\sim$0.44 meV/\AA$^2$. It is worth noting that, within the
current computational resources, the variable-$n$ algorithm
performs quite well even when the number of atoms is doubled. As a
test, we have run a variable-$n$ calculation with all the
structures in the pool having a number of atoms ($n=406$) which
does not correspond to the optimal configuration: this simulation
still retrieves the SR model (global minimum), within about $10^4$
genetic operations.

In conclusion, we have shown that the reconstruction of
semiconductor surfaces can be determined via a genetic algorithm.
This procedure can be used to generate a database of model
configurations for any given high-index surface, models that can
be subsequently relaxed using electronic structure methods and
compared with available experimental data. The process of
systematically building a set of models for a given surface
drastically reduces the probability of missing the actual physical
reconstruction, which imminently appears when heuristic approaches
are used (see, e.g., Refs. \cite{5512old, mo}). The genetic
algorithm presented here can naturally select the number of atoms
required for the topology of the most stable reconstruction, as
well as the lowest-energy bonding configuration for that number of
atoms. Future work will be focused on other high-index silicon and
germanium surfaces \cite{5512new, si-highindex, ge-highindex}, and
on the structure of clusters deposited on silicon surfaces
\cite{clusteronsurface}.

{\bf Acknowledgments.} Ames Laboratory is operated for the U.S.
Department of Energy by Iowa State University under Contract No.
W-7405-Eng-82; this work was supported by the Director of Energy
Research, Office of Basic Energy Sciences. Support from the MRSEC
at Brown University (DMR-0079964), NSF (CMS-0093714, CMS-0210095),
and the Salomon Research Award from the Graduate School at Brown
University is gratefully acknowledged.

\newpage
\newpage
\newpage
\newpage
\newpage
\begin{table}
\begin{center}
\begin{tabular}
{ccc} \hline\hline
$n$ & Surface energy   & Label \\
    & (meV/\AA $^{2}$) & from Refs.~\cite{ptmc, susc105} \\
\hline
206 & 82.20 & SR  \\
    & 85.12 & DT  \\
    & 88.12 & DU1 \\
    & 88.28 &     \\
    & 88.35 & SU  \\
205 & 86.73 & DY1 \\
    & 88.59 & DY2 \\
    & 88.61 &     \\
    & 88.70 &     \\
    & 88.97 &     \\
204 & 84.90 & DX1 \\
    & 86.04 & DX2 \\
    & 88.98 &     \\
    & 89.11 &     \\
    & 89.78 &     \\
203 & 86.52 & DR1 \\
    & 87.74 &     \\
    & 89.49 &     \\
    & 90.50 &     \\
    & 90.54 &     \\
\hline\hline
\end{tabular}
\end{center}
\caption{Surface energies of 20 different Si(105) reconstructions
obtained by the genetic algorithm (GA), calculated using the HOEP
interatomic potential \cite{hoep}. The structures are grouped
according to the number of atoms $n$ in the simulation cell. The
ground states are the same as those reported in Ref. \cite{ptmc}.}
\label{tabelul}
\end{table}

\end{document}